\documentclass[sigconf,screen,authorversion]{acmart}

\usepackage{balance}
\usepackage{url}
\usepackage{amsmath,amsfonts}
\usepackage{algorithmic}
\usepackage{graphicx}
\usepackage{textcomp}
\usepackage{xcolor,colortbl}

\AtBeginDocument{%
  }

\setcopyright{acmlicensed}
\acmPrice{15.00}
\acmDOI{10.1145/3597926.3598112}
\acmYear{2023}
\copyrightyear{2023}
\acmSubmissionID{issta23main-p271-p}
\acmISBN{979-8-4007-0221-1/23/07}
\acmConference[ISSTA '23]{Proceedings of the 32nd ACM SIGSOFT International Symposium on Software Testing and Analysis}{July 17--21, 2023}{Seattle, WA, United States}
\acmBooktitle{Proceedings of the 32nd ACM SIGSOFT International Symposium on Software Testing and Analysis (ISSTA '23), July 17--21, 2023, Seattle, WA, United States}
\received{2023-02-16}
\received[accepted]{2023-05-03}

\usepackage{enumitem}
\usepackage{tikz}
\usepackage{amsmath}

\newcommand\ChangeRT[1]{\noalign{\hrule height #1}}
\usepackage{multirow}
\usepackage{makecell}
\usepackage[linesnumbered,ruled,vlined]{algorithm2e}
\SetKwComment{Comment}{$\triangleright$\ }{}
\SetKwComment{LeftComment}{$\triangleright$ \hfill~}{}

\usepackage{comment}
\usepackage{doi}

\newcommand{\wvar}{\textsc{$\omega$Var}}
\newcommand{\wvarws}{\textsc{$\omega$Var }}

\newcommand{\wTester}{\textsc{$\omega$Test}}
\newcommand{\wTesterws}{\textsc{$\omega$Test }}
\newcommand{\wAPI}{\textsc{$\omega$Test-API}}
\newcommand{\wAPIws}{\textsc{$\omega$Test-API }}
\newcommand{\wBug}{\textsc{$\omega$Bug}}

\newcommand{\wDroid}{\textsc{$\omega$Droid}}
\newcommand{\wDroidws}{\textsc{$\omega$Droid }}

\newcommand{\osappsws}{44 }

\newcommand{\csappsws}{30 }

\newcommand{\allappsws}{74 }

\newcommand{\detectedBugsws}{36 }

\newcommand{\submittedBugsws}{22 }

\newcommand{\confirmedBugsws}{13 }

\newcommand{\fixedBugsws}{9 }

\definecolor{Gray}{gray}{0.9}

\newcolumntype{a}{>{\columncolor{Gray}}p{1mm}}

\begin{document}

\title{\wTester: WebView-Oriented Testing for Android Applications}

\author{Jiajun Hu}
\email{jhuao@cse.ust.hk}
\affiliation{%
	\institution{The Hong Kong University of Science and Technology}
	\city{Hong Kong}
	\country{China}
}

\author{Lili Wei}
\email{lili.wei@mcgill.ca}
\affiliation{%
	\institution{McGill University}
	\city{Montreal}
	\country{Canada}
}

\author{Yepang Liu}
\authornote{Yepang Liu is affiliated with both the Department of Computer Science and Engineering and the Research Institute of Trustworthy Autonoumous Systems, Southern University of Science and Technology.}
\email{liuyp1@sustech.edu.cn}
\affiliation{%
	\institution{Southern University of Science and Technology}
	\city{Shenzhen}
	\country{China}
}

\author{Shing-Chi Cheung}
\authornote{Corresponding Author}
\email{scc@cse.ust.hk}
\affiliation{%
	\institution{The Hong Kong University of Science and Technology}
	\city{Hong Kong}
	\country{China}
}


\begin{abstract}
  WebView is a UI widget that helps integrate web applications into the native context of Android apps. It provides powerful mechanisms for bi-directional interactions between the native-end (Java) and the web-end (JavaScript) of an Android app. However, these interaction mechanisms are complicated and have induced various types of bugs. To mitigate the problem, various techniques have been proposed to detect WebView-induced bugs via dynamic analysis, which heavily relies on executing tests to explore WebView behaviors. Unfortunately, these techniques either require manual effort or adopt random test generation approaches, which are not able to effectively explore diverse WebView behaviors. In this paper, we study the problem of test generation for WebViews in Android apps. Effective test generation for WebViews requires identifying the essential program properties to be covered by the generated tests. To this end, we propose \emph{WebView-specific properties} to characterize WebView behaviors, and devise a cross-language dynamic analysis method to identify these properties. We develop \wTester, a test generation technique that searches for event sequences covering the identified WebView-specific properties. An evaluation on \allappsws real-world open-/closed-source Android apps shows that \wTesterws can cover diverse WebView behaviors and detect WebView-induced bugs effectively. \wTesterws detected \detectedBugsws previously-unknown bugs. From the \submittedBugsws bugs that we have reported to the app developers, \confirmedBugsws bugs were confirmed, \fixedBugsws of which were fixed.
\end{abstract}



\maketitle

\section{Introduction}\label{introduction}
WebViews are user interface (UI) widgets of the Android framework to display web pages in Android apps~\cite{WebView}. They are instances of the \texttt{android.webkit.WebView} class.
WebViews not only help integrate web applications developed in HTML/JavaScript into Android apps but also provide mechanisms to support interactions between the web-end (JavaScript) and the native-end (Java) of an Android app.
WebViews are widely used in real-world Android apps. An earlier study~\cite{mutchler2015large}, which analyzed over one million apps from Google Play store, showed that 85\% apps use WebViews in some fashion. A recent study~\cite{zhang2022identity} that randomly crawled 6000 popular apps from two leading app stores demonstrated similar results (i.e, 90.6\% of them use WebViews). A dataset containing 6400 most-popular Google Play apps~\cite{AppBrain1} recently crawled by us also showed that 83.74\% apps use WebViews in their code.
Furthermore, a new trending ecosystem called app-in-app (e.g., WeChat Mini-Programs~\cite{miniprogram}) in which a host-app uses WebViews to embed sub-apps is adopted by many high-profile apps~\cite{lu2020demystifying, wang2022wechat, zhang2021measurement, zhang2022identity}. In reality, millions of active users are interacting with WebViews everyday~\cite{wechatusers}.

Despite the popularity, WebView programming is error-prone due to its complicated cross-language interaction mechanisms (e.g., bridge communication~\cite{lee2016hybridroid, bae2019towards} and WebView callbacks~\cite{yang2018automated}).
Existing studies have investigated various types of bugs induced by the misuses of WebViews (e.g., security issues~\cite{li2017unleashing, lee2016hybridroid, rizzo2018babelview, luo2011attacks, song2018understanding, bai2018bridgetaint, tang2018dual, yang2018automated, yang2018study, yang2019iframes, zhang2018empirical,el2021vulnerabilities}). They also proposed different bug detection techniques based on static/dynamic analysis. A fundamental limitation of static analysis techniques~\cite{rizzo2018babelview, song2018understanding, lee2016hybridroid, yang2019iframes,zhang2018empirical,el2021vulnerabilities} is their inability of analyzing dynamically loaded web data. For example, some JavaScript code may not be available for analysis until they are loaded at runtime. In comparison, dynamic analysis techniques~\cite{bai2018bridgetaint, hu2018tale, tang2018dual, li2017unleashing, yang2018automated, yang2018study} exercise WebViews through testing, and therefore do not suffer from such a limitation.
Nonetheless, the effectiveness of dynamic analysis techniques heavily relies on the quality of the generated tests.
For example, manifesting a WebView-induced bug
may require the tests to trigger some specific events on the buggy WebViews. Yet, most of the existing dynamic analysis techniques either adopt manual~\cite{bai2018bridgetaint} or random approaches for test generation~\cite{hu2018tale, yang2018study, yang2018automated}. Their generated tests cannot effectively explore diverse WebView behaviors to expose hidden issues.
Conventional general-purpose test generation techniques for Android apps are also ineffective for WebView testing.
A typical category of these techniques is model-based graphical user interface (GUI) testing~\cite{su2017guided, baek2016automated, gu2019practical, pan2020reinforcement, dong2020time, degott2019learning,wang2021vet,fastbot2}, whose objective is to generate tests to visit more GUI states of an Android app. The GUI state of an app, which is modeled by the hierarchy of the rendered UI elements, is an abstraction of app behaviors. Intuitively, visiting more GUI states that ``look different'' means that more app behaviors are explored.
However, GUI states may not be a good abstraction of WebView behaviors. Loading a new page in a WebView does not necessarily mean that a new WebView behavior is explored. For example, opening two different websites in a WebView-wrapped browser app may exercise the same page-loading process of WebViews.
Another typical category of conventional techniques is to guide test generation by the coverage of some program properties~\cite{su2017guided,mao2016sapienz,mahmood2014evodroid,wang2020combodroid} that reflect the test objectives. Program statements and branches are two commonly adopted properties.
For example, Sapienz~\cite{mao2016sapienz}, an Android test generation technique, regards the maximization of statement coverage as a main objective in test generation.
However, the program properties adopted by existing work are not suitable for testing WebViews since none of them are specifically designed to model WebView behaviors. For example, not all statements in an app are used for implementing WebView functions. Leveraging these properties cannot guide the generation of tests to systematically examine WebView behaviors.
Therefore, an effective test generation technique targeting WebViews in Android apps is needed.

To effectively test WebViews, it is desirable to define specific test objectives and propose properties accordingly to guide test generation.
A straightforward solution is to take WebView API\footnote{WebView APIs include APIs of classes under packages \texttt{android.webkit} and \texttt{androidx.webkit}, and bridge methods~\cite{webApps}.}
call sites as the properties for tests to cover.
However, such a design of properties is inadequate: it only captures the interaction sites between the web-end and native-end but ignores the involved data exchanges.
As we will illustrate in Section~\ref{motivation}, effective WebView testing should also examine how the data sent from the web-end are manipulated at the native-end and vice versa.
Based on this observation, we propose a novel design of \emph{WebView-specific properties} that considers both WebView API call sites and the data exchanges involved in the web-native interactions.
We also devise a cross-language dynamic analysis method to identify these properties.
We further propose a WebView-oriented test generation technique \wTester. \wTesterws is powered by a novel fitness function, which awards those tests that can cover diverse WebView-specific properties.


We evaluated \wTesterws on \osappsws open-source and \csappsws closed-source Android apps and compared it with 5 baseline methods in terms of property coverage and the number of detected WebView-induced bugs. 
Our evaluation results show that \wTesterws achieved the highest coverage and detected the most number of bugs.
 \wTesterws detected \detectedBugsws previously-unknown bugs. From the \submittedBugsws bugs that we have reported to the corresponding app developers,
\confirmedBugsws bugs were confirmed and \fixedBugsws of them have been fixed.
 To summarize, this paper makes the following major contributions:

\begin{itemize}[label=\textbullet, leftmargin=*, itemsep=3pt]
	\item \textbf{WebView-specific properties:} We propose the first design of WebView-specific properties to formally characterize WebView behaviors in Android apps.
	\item \textbf{\wTester:} We propose a test generation technique \wTester, which leverages the proposed properties to generate tests that are able to explore diverse WebView behaviors.
	\item \textbf{Implementation and evaluation:} We implemented \wTesterws and evaluated its performance on real-world Android apps.
	It detected \detectedBugsws previously-unknown bugs in various apps. We make the tool and experiment dataset available to facilitate future research~\cite{wTest}.
\end{itemize}


\section{Motivation and Property Design}\label{motivation}

\begin{figure*}
	\centering
	\includegraphics[width=1.0\linewidth]{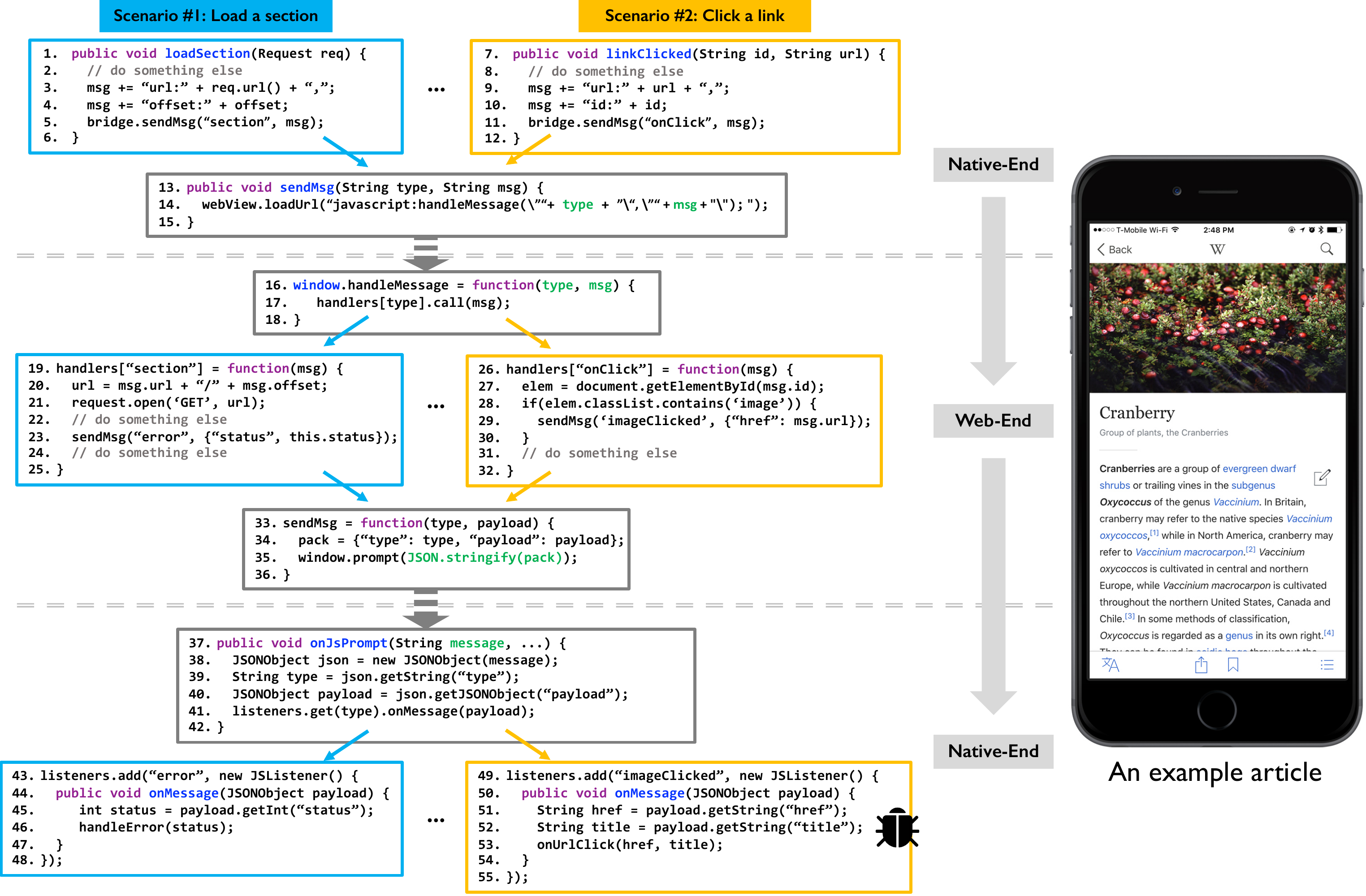}
	\caption{Two web-native interaction scenarios in the Wikipedia Android app (the code is simplified and adapted to ease understanding).}
	\label{fig:example}
\end{figure*}

Program properties (e.g.,  program statements) are entities that characterize the program behaviors of interest. The coverage of these properties provides an effective measure of test adequacy for the interested behaviors. We consider both WebView APIs call sites and the data exchanges in the web-native interactions as \emph{WebView-specific properties}.
We demonstrate the need of doing so using two web-native interaction scenarios in Wikipedia~\cite{Wikipedia}. 


\subsection{Two Interaction Scenarios}
Figure~\ref{fig:example} gives the code of the two interaction scenarios. Boxes and arrows marked in blue and yellow represent Scenario~\#1 and \#2, respectively. Boxes marked in grey represent the functions shared by both scenarios.
Both of the scenarios (1) are initiated at the native-end, (2) invoke a JavaScript function to pass data to the web-end via the WebView API \texttt{loadUrl()} (line~14), and (3) send the results generated at the web-end back to the native-end via the WebView callback \texttt{onJsPrompt()} (line~37). The two scenarios share the same interfaces that handle cross-language data transmissions (grey boxes).
The handling is different according to the interaction \emph{type}.
Scenario~\#2 contains a reported bug~\cite{T149692}.

Scenario~\#1 depicts the process of loading an article section. Its interaction type is ``\emph{section}''.
Figure~\ref{fig:example} shows a section introducing cranberry.
In this scenario, the native-end constructs a JSON-format \texttt{msg} containing a requested \texttt{url} (line~3) and an \texttt{offset} (line~4). It calls \texttt{loadUrl()} to pass \texttt{msg} and the interaction \texttt{type} (\texttt{"section"} specified at line 5) as arguments to the web-end through a dynamically constructed JavaScript code (line 14).
The web-end retrieves the data in \texttt{handleMessage()} and dispatches \texttt{msg} to the handler for type \texttt{"section"} (line 17).
The handler (lines 19\textendash25) then fetches the resources through an HTTP request based on the information in \texttt{msg} (lines 20\textendash21).
If an error occurs, the error status will be sent back to the native-end (line 23).
The error message is sent by prompting a dialog (line 35) and thereby triggering the native-end WebView callback \texttt{onJsPrompt()} (lines 37\textendash42).
The native-end receives the error as the callback parameter (line 37) and handles it at line 46.

Scenario~\#2 depicts the process of handling a click on a link. Its interaction type is ``\emph{onClick}''.
This scenario is initiated at the native-end (lines 7\textendash12) by packing the link \texttt{url} and the \texttt{id} of the DOM element that consumes the click event into \texttt{msg} (lines 9\textendash10).
It calls \texttt{loadUrl()} (line 14) and passes the \texttt{msg} to the handler (line 26) for \texttt{"onClick"} at the web-end.
If the DOM element contains an image (line 28), a message of type \texttt{"imageClicked"} is sent back to the corresponding event handler at the native-end (line 49).
This scenario contains a bug that results in a crash when a \texttt{JSONException} is thrown at line 52~\cite{T149692}. The exception occurs because the Java code is looking for a \texttt{"title"} attribute from a JSON-typed object \texttt{payload} (line 52) prepared by the JavaScript code at line 29, where the \texttt{"title"} attribute is not set.

\subsection{Limitation of WebView API Call Sites}
We can see from the example that it is inadequate to characterize different web-native interaction scenarios by considering only WebView API call sites (\texttt{loadUrl()} at line 14 and \texttt{onJsPrompt()} at line 37) as properties for test coverage.
While the two scenarios explore different WebView behaviors, a test enacting either one of them can cover all these call sites.
A test generation technique solely driven by WebView API call sites may consider the test exploring Scenario \#2 redundant after generating the test for Scenario \#1, thus missing the bug residing in Scenario \#2.
This motivates us to propose new properties that can better characterize WebView behaviors.
\subsection{New Design of WebView-Specific Properties}
From the example, we can make a key observation. Besides the call sites of WebView APIs, it is necessary to consider the data exchanges across the language boundaries to characterize WebView behaviors.
In other words, \textbf{\textit{program variables that carry the transmitted data between the web-end and the native-end should be captured by WebView-specific properties.}} We call such variables \wvar s. 
A higher coverage of these properties is likely to explore more WebView behaviors. For example, a test that covers \wvar s \texttt{req.url()}, \texttt{offset}, and \texttt{status} exercises Scenario \#1, while a test that covers \wvar s \texttt{href}, \texttt{id}, and \texttt{title} exercises Scenario \#2.
With \wvar s, two scenarios can be distinguished from each other.
However, it is challenging to identify \wvar s.
Unlike conventional program properties such as program statements or branches, \wvar s cannot be simply identified from program structures. 
Their identification involves two challenges:

\textbf{Challenge 1:} JavaScript code can be dynamically constructed at runtime. If it cannot be precisely determined, we may miss \wvar s at the web-end. For example, line 14 in Figure~\ref{fig:example} builds a string of JavaScript code. When a section is loaded, the string can be \emph{javascript:handleMessage(``section'', \{url: ``...'', offset: ...\});}. The \texttt{url} and \texttt{offset} field of the object expression (the second argument of \texttt{handleMessage()}) should be identified as \wvar s since their values are set by the Java code. Without recognizing the JavaScript code, these two \wvar s can be missed, further affecting the identification of more \wvar s that depend on them (e.g., \texttt{url} at line 20).


\textbf{Challenge 2:} Not all variables are \wvar s. 
It is difficult to statically identify \wvar s precisely due to language or Android framework features (e.g., dynamic data structures, intent, etc.) that involve dynamic decisions.
For example, static analysis often adopts an over-approximation strategy when analyzing collections. It may consider all the elements in an array to be \wvar s even though only some of them carry transmitted data.

If these \wvar s cannot be precisely identified, the generated tests are ineffective. To address the challenges, we devise a set of \textit{dynamic} \wvarws tagging rules to identify WebView-specific properties \textit{on-the-fly} during testing.
We present the method in the next section.

\section{Dynamically Identifying WebView-Specific Properties}\label{property}

\begin{figure}
	\centering
	\includegraphics[width=1.0\linewidth]{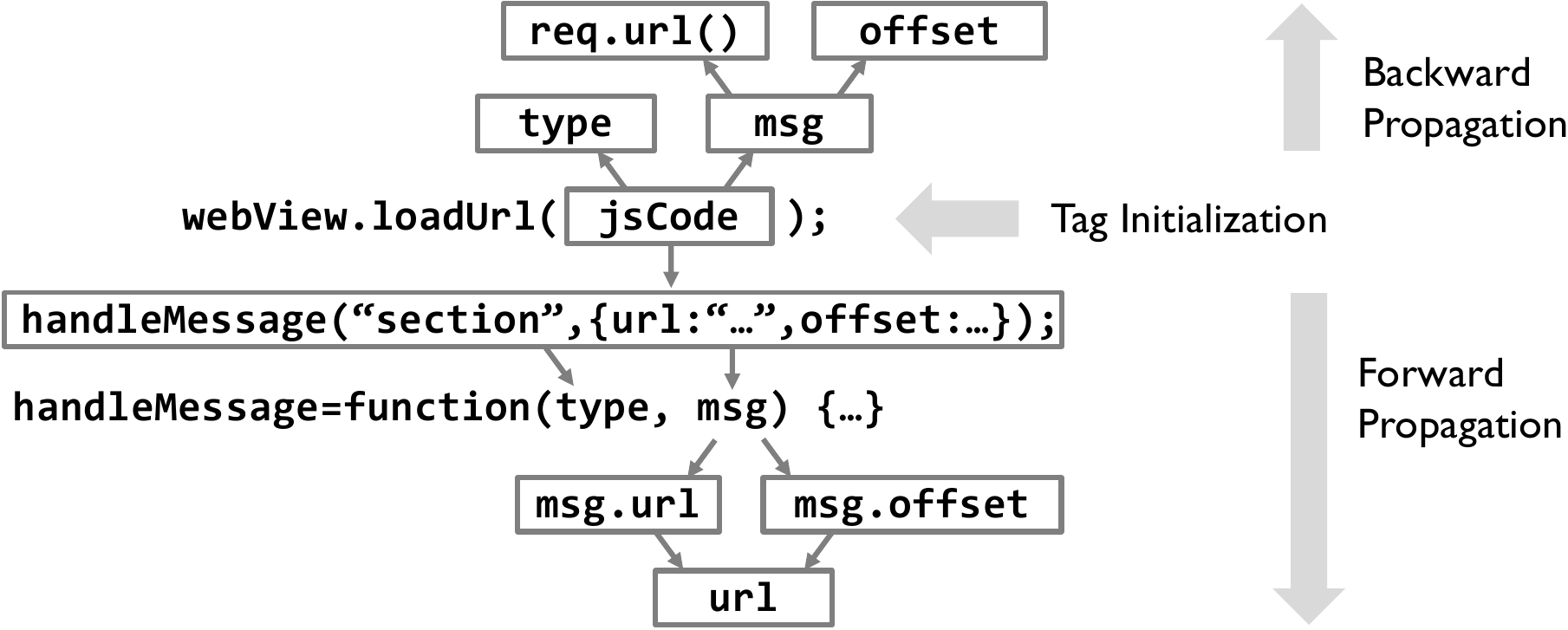}
	\caption{An overview of \wvarws identification.}
	\label{fig:propOverview}
\end{figure}

This section presents a dynamic analysis method to identify WebView-specific properties in an Android app.
Figure~\ref{fig:propOverview} gives an overview of the method.
The idea is to first initialize a set of \wvar s that are directly used by WebView APIs (\emph{\textbf{Tag Initialization}}).
Then, other variables depending on existing \wvar s and vice versa are iteratively identified and
added to the set of \wvar s  using \emph{\textbf{Forward}} and \emph{\textbf{Backward Propagation}}, respectively.
Figure~\ref{fig:propOverview} abstracts the dependencies between variables in Scenario~\#1 in Figure~\ref{fig:example}.
In this example, when \texttt{loadUrl()} is executed by a test, its argument \texttt{jsCode} is the first variable to be tagged as an \wvar. Then the variables that \texttt{jsCode} depends on are iteratively tagged as \wvar s by traversing the variable dependencies built during test execution according to a set of \emph{backward propagation} rules. For example, since \texttt{jsCode} is built from \texttt{type} and \texttt{msg}, both of them are tagged as \wvar s.
Variable \texttt{msg} further depends on the return value of \texttt{req.url()}\footnote{We instrument an app to inject the propagation logic at instruction level. There will be a temporary variable to hold the return of req.url() and this variable will be tagged.} and \texttt{offset}, the corresponding variables are also tagged as \wvar s.
Next, \texttt{loadUrl()} will execute the JavaScript code.
We analyze the code and identify the JavaScript variables whose values are defined by Java code.
In this example,
the string argument \texttt{"section"}, the \texttt{url} and \texttt{offset} field of the object expression (the second argument) of \texttt{handleMessage()} are tagged as \wvar s.
We further use a set of \emph{forward propagation} rules to tag new \wvar s depending on existing \wvar s at the web-end.
In Figure~\ref{fig:propOverview}, 
\texttt{url} is tagged since it depends on two existing \wvar s,  \texttt{msg.url} and \texttt{msg.offset}.

Table~\ref{table:propagation} shows the tagging rules to identify \wvar s.
We use a function $\omega(x)$ to indicate if a variable $x$ is an \wvarws: if $x$ is an \wvar, the value of $\omega(x)$ is 1; otherwise 0. The code that implements the tagging logic is injected into an app using program instrumentation. Therefore, it is independent from any test generation approaches.
In the following, we present the three steps to identify \wvar s: \textit{Tag Initialization, Backward Propagation, and Forward Propagation}.


\begin{figure}
	\centering
	\includegraphics[width=0.5\linewidth]{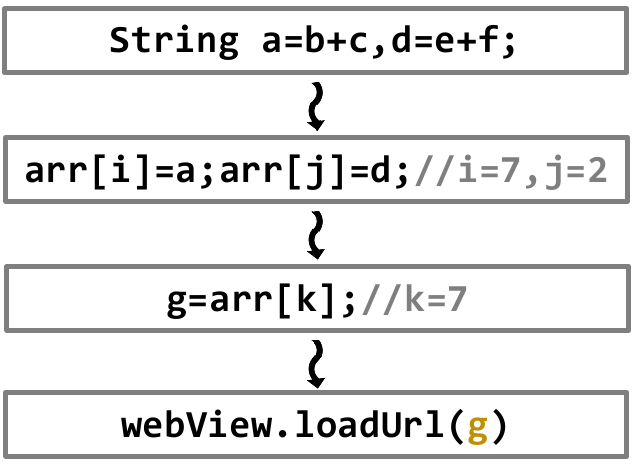}
	\caption{Motivation of variable tracking}
	\label{fig:varTracking}
\end{figure}

\subsection{Tag Initialization}
In the first step, we identify an initial set of \wvar s based on \textit{tag initialization rules}, which consider the following five circumstances according to the semantics of WebView APIs~\cite{webApps}. 

(1) \emph{Parameters and return variables of WebView callback methods are tagged as  \wvar s}. These native-end parameters carry the information of events occurring at the web-end. For example, the parameter \texttt{message} of the callback method \texttt{onJsPrompt()} at line 37 in Figure~\ref{fig:example} carries the prompted message sent by \texttt{window.prompt()} at line 35. The callback return also affects WebView behaviors. For instance, WebViews determine whether to continue loading a URL according to the return of \texttt{shouldOverrideUrlLoading()} when a user clicks a URL.

\begin{table*}
	\footnotesize
	\begin{center}
		\caption{\wvarws Tagging Rules ([] means optional)}
		\begin{tabular}{| c | c | c | c | c |}
			\ChangeRT{1pt}
			\textbf{} & \textbf{No.} & \textbf{Operation Semantics} & \textbf{Propagation/Tagging Rules} & \textbf{Applied Language} \\
			\ChangeRT{1pt}
			\hline
			\multirow{5}{*}{\makecell{Tag \\ Initialization}} & 1 & $webview\_callback([y_1, y_2, ..., y_n]) \{...[return\;r;]\}$ & $\omega(y_1)=1, \omega(y_2)=1, ..., \omega(y_n)=1, \omega(r) = 1$ & Java\\
			\cline{2-5}
			& 2 & \makecell{$bridge\_mtd([y_1, y_2, ..., y_n]) \{... [return\;r;]\}$} & $\omega(y_1)=1, \omega(y_2)=1, ..., \omega(y_n)=1, \omega(r) = 1$ & Java\\
			\cline{2-5}
			&3 & $[x=] webview\_api\_Ivk([y_1, y_2,...,y_n])$ & $\omega(y_1)=1, \omega(y_2)=1, ..., \omega(y_n)=1, \omega(x) = 1$ & Java\\
			\cline{2-5}
			& 4 & $[x =] bridge\_ivk([y_1, y_2, ..., y_n])$ &$\omega(y_1)=1, \omega(y_2)=1, ..., \omega(y_n)=1, \omega(x) = 1$ & JavaScript\\
			\cline{2-5}
			& 5 & $execJS(jsCode)$& $\{\omega(x)=1 | x\in jsCode \wedge isLiteral(x) == true\}$ & JavaScript\\
			\ChangeRT{1pt}
			\hline
			\multirow{5}{*}{\makecell{Backward \\ Propagation}} & 6 & $x = op\;y$ & $\omega(x) \rightarrow \omega(y)$ & Java \& JavaScript\\
			\cline{2-5}
			& 7 & $x = y\;op\;z$ & \makecell{$\omega(x) \rightarrow \omega(y), \omega(x) \rightarrow \omega(z)$} & Java \& JavaScript\\
			\cline{2-5}
			& 8 & $x = lib(y_1, y_2, ..., y_n)$\textsuperscript{1} & \makecell{$\omega(x) \rightarrow \omega(y_1), \omega(x) \rightarrow \omega(y_2), ..., \omega(x) \rightarrow \omega(y_n)$}  & Java \& JavaScript\\
			\cline{2-5}
			& 9 & $x.y = z$ & $\omega(x) \rightarrow \omega(z)$ & Java \& JavaScript\\
			\cline{2-5}
			& 10 & $arr[i] = x$ & $\omega(arr) \rightarrow \omega(x)$  & Java \& JavaScript\\
			\ChangeRT{1pt}
			\hline
			\multirow{5}{*}{\makecell{Forward \\ Propagation}} & 11 & $x = op \;y$ & $\omega(y) \rightarrow \omega(x)$ & Java \& JavaScript\\
			\cline{2-5}
			& 12 & $x = y\;op\;z$ & $\omega(y) | \omega(z) \rightarrow \omega(x)$ & Java \& JavaScript\\
			\cline{2-5}
			& 13 & $x = lib(y_1, y_2, ..., y_n)$\textsuperscript{1} & $\omega(y_1) | \omega(y_2) | ... | \omega(y_n) \rightarrow \omega(x)$ & Java \& JavaScript\\
			\cline{2-5}
			& 14 & $x = y.z$ & $\omega(y.z) \rightarrow \omega(x)$ & Java \& JavaScript\\
			\cline{2-5}
			& 15 & $x = arr[i]$ & $\omega(arr[i]) \rightarrow \omega(x)$ & Java \& JavaScript \\
			\ChangeRT{1pt}
			\multicolumn{5}{l}{\textsuperscript{1}lib stands for library methods, which include Android system APIs and APIs in third party libraries. We rely on existing tools~\cite{ma2016libradar,backes2016reliable} to identify them if an app's source code} \\
			\multicolumn{5}{l}{\textsuperscript{ }is not available (i.e., closed-source apps).}
		\end{tabular}
		\label{table:propagation}
	\end{center}
\end{table*}

(2) \emph{Parameters and return variables of bridge methods~\cite{lee2016hybridroid} are tagged as  \wvar s}. Bridge methods are Java methods that can be directly invoked by JavaScript~\cite{webApps}. The parameters of bridge methods are passed from the web-end to the native-end. The result computed at the native-end is passed back to the JavaScript code by the return values of bridge methods.

(3) \emph{Arguments passed to WebView APIs and variables saving return values of WebView API invocations are tagged as \wvar s}. These APIs are invoked to control the behaviors of WebViews (e.g., the display of a web page) through their arguments or extract information from the web-end (e.g., the current URL) through their return values.

(4) \emph{Arguments passed to bridge method invocations and the variables saving the return values of bridge method invocations are tagged as \wvar s}.
Different from rule (2), this rule captures the variables involved with bridge methods at the web-end.

(5) \emph{The literals in the JavaScript code that is dynamically constructed and loaded by WebView APIs that can execute JavaScript (e.g., \texttt{loadUrl()}) are tagged as \wvar s}.
The reason for tagging only the literals is that once a string representing JavaScript code is formatted, only the values of literals are given by the Java code. For example, the constructed JavaScript code at line 14 of Figure~\ref{fig:example} can be \emph{javascript:handleMessage(``section'', \{url: ``https://...'', offset: 3\})} when sections are loaded. The literals \emph{``section''}, \emph{``https://...''}, and \emph{3} are tagged as \wvar s as their values are given by three Java variables, respectively (i.e., \texttt{type}, \texttt{req.url()}, \texttt{offset}). We implement this by intercepting the constructed JavaScript code at runtime. We instrument the JavaScript code to inject the tagging logic into it before the code is loaded by WebView APIs.

\subsection{Backward Propagation}
Backward propagation is used to iteratively identify new \wvar s that existing \wvar s depend on. Backward propagation based on static analysis may misclassify many variables as \wvar s. 
Figure~\ref{fig:varTracking} depicts one scenario.  At the beginning, two string variables \texttt{a} and \texttt{d} are put into \texttt{arr[i]} (i=7) and \texttt{arr[j]} (j=2), respectively. Then at another place \texttt{arr[k]} (k=7) is assigned to \texttt{g}, which will be tagged as an \wvarws when it is used by \texttt{loadUrl()}. In this scenario, it is difficult to statically infer that only \texttt{arr[7]} is an \wvar. An over-approximation strategy based on static analysis can incorrectly classify all the elements in the array as \wvar s and subsequently tag properties irrelevant to WebView behavior. In this case, both \texttt{a} and \texttt{d} will be tagged as \wvar s, which will further lead to the tagging of \texttt{b}, \texttt{c} (which \texttt{a} depends on) and \texttt{e}, \texttt{f} (which \texttt{d} depends on). However, only \texttt{a} is used by \texttt{loadUrl()}.
As such, we address the problem using a dynamic \emph{\textbf{variable tracking}} strategy.

\subsubsection{Variable Tracking}
We adapt the idea of Phosphor~\cite{bell2014phosphor, bell2015dynamic} and BridgeTaint~\cite{bai2018bridgetaint} to track variables in Java and JavaScript respectively at runtime. The idea is to leverage a unique \emph{ID} assigned to each variable at its initialization to precisely identify created variables at runtime. 
During an app's execution, variable dependencies specified by the backward propagation rules are memorized using the ID assigned to each variable. When a WebView API is invoked, we will use the recorded dependencies to propagate.
Take Figure~\ref{fig:varTracking} as an example. Suppose that the IDs of \texttt{a}, \texttt{b}, \texttt{c} are 3, 1, 2, respectively. We can construct a data dependency that is $1, 2 \Rightarrow 3$ when executing \texttt{a=b+c;}. When \texttt{loadUrl()} is executed, we will look at the ID of \texttt{g}, which is 3 because \texttt{g} is a reference of \texttt{a}, and tag it as an \wvar. Then the backward propagation will be triggered to search for IDs that 3 depends on, which are 1 and 2, and tag the corresponding variables as \wvar s (i.e., \texttt{b} and \texttt{c}). The over-approximation problem can be much alleviated this way.

\subsubsection{Backward Propagation Rules}
The backward propagation rules are presented in the \emph{Backward Propagation} section of Table~\ref{table:propagation}. 
The five rules identify new \wvar s that provide values to existing \wvar s in assignment expressions.

\subsection{Forward Propagation}
The five forward propagation rules identify new \wvar s that depend on existing \wvar s in assignment expressions.
Unlike backward propagation which relies on dependencies memorized during app execution, forward propagation is immediately built whenever a program statement that matches one of the rules is executed. For example, when $x = y\ op\ z$ is executed and one of $y$ or $z$ is an \wvar, $x$ will be directly tagged as a new \wvarws according to rule No.12. We also make forward propagation field-sensitive, i.e., when a newly created \wvarws is a reference to an object/array/collection, its fields/elements will be recursively tagged as \wvar s.


\subsection{WebView-Specific Properties}
Finally, we summarize the set of identified WebView-specific properties. At runtime, all variables including \wvar s are uniquely represented by their assigned IDs.
It is a key step to alleviate the over-classification problem of \wvar s.
However, the assigned IDs cannot be directly treated as the properties that tests should cover.  It is because IDs are assigned in a non-deterministic way at runtime (i.e., it is increased  by 1 whenever a new variable is initialized) so that even two identical tests may generate two different sets of IDs. Therefore, tests cannot be measured by the covered ID set. To generate a deterministic property set, we propose to take the \emph{def} locations of \wvar s ($Def_\wvar$) found by backward propagation and the \emph{use} locations of \wvar s ($Use_\wvar$) found by forward propagation as the property set. The \emph{def} location of an \wvarws is the place in a program where the ID of that \wvarws is assigned or the place where a variable dependency on that \wvarws is built. The \emph{use} location of an \wvarws is the place in a program where that \wvarws is used (e.g., it is used as an argument of a method invocation). These locations are also uniquely indexed. To summarize, the WebView-specific property set that a test covers ($P$) is the union of the covered WebView API call sites ($\omega API$), $Def_\wvar$, and $Use_\wvar$:
\begin{equation} \label{propertySet}
P = \omega API \cup Def_\wvar \cup Use_\wvar
\end{equation}

\subsection{Implementation}\label{instru_imple}
We implement the tagging/propagation logic by app instrumentation. We use Soot~\cite{soot} and Esprima~\cite{esprima} to instrument all the Java and JavaScript statements that match the operation semantics in Table~\ref{table:propagation}, respectively. We only instrument the JavaScript code located inside the asset folder of an app (which can be accessed by decoding an apk file via Apktool~\cite{apktool}) and the JavaScript code dynamically loaded by WebView APIs (e.g., the one loaded by \texttt{loadUrl()}). Other JavaScipt code (e.g., code in online websites) are not instrumented as they usually do not interact with the native-end of an app. The properties covered in the JavaScript code are sent to the native-end via bridge communication~\cite{webApps}. Together with the properties covered in the Java code, they are stored in the memory and will be saved to a file in the hard disk of an Android emulator or a real device every 500 ms. The file can be read by any test generation tools to retrieve coverage information.
To also measure traditional coverage such as statement/method coverage, we also implement the logic of JaCoCo~\cite{JaCoCo} into our instrumentation tool. The tool supports statement/method coverage on both Java and JavaScript code. To ease the variable tracking, we require the instrumented app to run on a customized Android OS. The tool and the OS are publicly available online~\cite{wTest}.

\section{Test Generation}\label{test}
This section presents how we utilize the identified WebView-specific properties to guide test generation for WebViews.
We implement the test generation procedure as a tool called \wTester, which aims to generate tests that maximize the coverage of WebView-specific properties of an Android app.

\subsection{Overview}\label{wTester}
\wTesterws is a search-based test generation technique that generates a sequence of events to optimize a fitness function designed to meet the objective above. Intuitively, it keeps appending events (e.g., click a widget) to exercise app UI components \footnote{UI components includes activities~\cite{Activity} and fragments~\cite{Fragment}.} if new properties are frequently discovered. It leaves a UI component if existing properties are repeatedly covered or no properties are found after enough events are tried.

\begin{algorithm}
	\caption{Test Generation Procedure (\wTester)}
	\label{alg:test}
	\KwIn{An app under test ($AUT$) and a time budget $Budget$}
	\KwOut{A set of covered WebView-specific properties $P$ and a set of WebView-induced bugs \wBug s}
	$P$ = $\varnothing$ \Comment*[r]{empty property set}
	\wBug s = $\varnothing$ \Comment*[r]{empty bug set}
	Launch $AUT$ \Comment*[r]{initial event}
	\While{$time \leqslant Budget$}
	{
		\If{$bug$ = foundBug()} {
			\wBug s = \wBug s $\cup$ $bug$ \Comment*[r]{find a bug}
		}
		\If{$P_{current}$ != $\varnothing$} {
			$P$ = $P$ $\cup$ $P_{current}$ \Comment*[r]{merge covered properties}
		}
		\eIf(\Comment*[f]{ A: the foreground activity}){continue($A$)} {
			\eIf{continue($A_F$) $||$ $|A|\leqslant1$} {
				\LeftComment{$A_F$: current fragments-state in A}
				\LeftComment{$|A|$: number of fragments-states in A}
				$e$ = selectAnEvent($UI$)\;
				executeEvent($e$)\;
			} {
			    switchFragmentsState()\;
			}
		} {
			pressBack()\;
		}
	}
\end{algorithm}

Algorithm~\ref{alg:test} depicts our proposed test generation procedure. It takes an app under test (AUT) that has been instrumented according to Section~\ref{property} and a time budget as inputs, and outputs a set of covered WebView-specific properties ($P$) and a set of discovered WebView-induced bugs ($\wBug s$). 
\wTesterws iteratively generates events to explore the AUT within the given time budget (lines 4--16).
It triggers an event in each iteration and monitors the manifested $\wBug s$ (lines 5--6) as well as 
the covered properties (lines 7--8).
\wTesterws decides whether to continually append new events (lines 10--14) or to leave the current activity $A$ (line 16) by utilizing a fitness function defined in $continue()$ (line 9). When the foreground activity has a good fitness value, \wTesterws will similarly compute a fitness value for the current fragments-state $A_F$ in $A$ (an activity can render multiple fragments on the screen, so the set of fragments currently displayed forms a fragments-state). When the fitness is good or there is zero or one fragments-state (i.e., $|A| \leqslant 1$, which means there is no more fragments-state to switch) in $A$, \wTesterws will randomly pick an available event (e.g., pressing a button, scrolling a list, inputting text, rotating screens, etc.) according to the current UI hierarchy (line 11) and execute it (line 12). Otherwise, an event that can switch to another fragments-state is triggered (line 14).
\wTesterws reports $\wBug s$ according to pre-defined test oracles (lines 5--6)~\cite{hu2018tale}.
In the following, we explain how \wTesterws
(1) decides whether to continue exploration from the current activity and fragments-state, and (2) identifies bugs with pre-defined test oracles.

\subsection{Continual Appending of Events}
\wTesterws determines whether to continue exploration according to a fitness value calculated to evaluate how good a state $S$ is. The state $S$ can be either an activity $A$ or a fragments-state ($A_F$) in $A$.  
In this section, we will (1) define the fitness function, and (2) explain how $continue()$ makes decision based on the fitness value.

\subsubsection{Fitness function}
\wTesterws aims to cover as many unique properties as possible while minimizing the number of times that a property is covered more than once in order to diversify the properties covered during test generation.
Test resources could be wasted if the generated tests keep visiting properties that have already been covered many times.
In addition, if a state's function is adequately explored by enough events but no WebView-specific properties are found, fewer test resources should be spent on that state.
Considering these factors and given a state $S$, the fitness function $f(S)$ is defined as:
\begin{equation} \label{objective}
f(S) = w \times f1(P_S, t_S) + (1-w) \times f2(N_S, c_S) 
\end{equation}
where $P_{S}$ is the set of properties covered when exploring $S$, $t_{S}$ is the number of times that new properties are found in $S$, $N_{S}$ is the number of events spent on $S$, $c_{S}$ is the number of times that code coverage increases in $S$, and $w$ is a weight that balances two sub-fitness functions $f1()$ and $f2()$.

$f1(P_{S}, t_{S})$ is inversely proportional to the frequency of the properties in $P_{S}$. Suppose $P_{S}$ is represented as $\{p_1, p_2, ..., p_m\}$ and the number of times that the properties in $P_{S}$ are covered is represented as $\{n_1, n_2, ..., n_m\}$. With this representation, $f1(P_{S}, t_{S})$ is defined as:
\begin{equation} \label{f1}
f1(P_S, t_S) = \frac{\sum_{i=1}^{m}{max(0, 1-(\frac{n_i / t_S}{\alpha})^\beta)}}{|P_S|}
\end{equation}

$f1()$ ranges between 0 and 1 and a higher value indicates a better fitness. The value of $f1()$ is high when new properties are frequently covered because in this case, the number of unique properties ($|P_{S}|$) and the number of times that property coverage increases ($t_{S}$) will be higher, and the frequency of each covered properties ($n_{i}$) will be relatively lower. The value of $f1()$ decreases when the covered properties are repeatedly covered (thus leading to higher $n_{i}$s). Two parameters $\alpha$ and $\beta$ control the decrease rate of $f1()$, which are set as $e$ and 2, respectively.



To effectively cover diverse WebView-specific properties in a limited time budget, we use $f2()$ to limit the number of events spent on a state to search for uncovered properties. We rely on the code coverage information to compute $f2()$. Intuitively, if no more code coverage increase is observed after enough events are spent on a state, that state should not waste any more test resources afterwards. With this design, $f2(N_{S}, c_{S})$ is defined as:
\begin{equation} \label{f2}
f2(N_S, c_S) = max(0, 1-(\frac{N_S / (c_S)^{r(c_S)}}{\epsilon})^\theta)
\end{equation}

Similarly, $f2()$ also ranges between 0 and 1. A higher value of $f2()$ indicates better fitness. When the number of times that code coverage increase observed in $S$ is high (i.e., a higher $c_{S}$), more events $N_{S}$ can be allocated to $S$ before $f2()$ returns a relatively small value. $r(c_{S})$ is a function that returns a value slightly smaller than 1. It is defined as $r(c_{S}) = 1-0.001*c_{S}$ to prevent the test from being stuck in a state whose code coverage is increased too many times. Two parameters $\epsilon$ and $\theta$ also control the decrease rate of $f2()$, which are set to 8 and 2, respectively.


When no properties are covered in a state, we only use $f2()$ to guide testing. When WebView-specific properties are covered in a state $S$, we make the fitness of $S$ (i.e., $f(S)$) biased to $f1()$ by setting the weight $w$ in Equation~\ref{objective} as 0.7. 

\subsubsection{Continue condition}
The function $continue()$ (line 9 \& line 10) makes decision based on the fitness value. It returns true if the following condition is satisfied: 

\begin{equation} \label{continueCondition}
Rand(0,1) < min(0.9, f(S))
\end{equation}
where $Rand(0,1)$ returns a random number between 0 (inclusive) and 1 (exclusive).
Intuitively, $f(S)$ can be treated as the probability that Condition~\ref{continueCondition} is satisfied. A higher $f(S)$ will make $continue()$ more likely to return true. Therefore, \wTesterws will have a higher probability to continue exploration from a state with good fitness. 

\subsection{Oracle}\label{oracle}
Test oracles (line 5) are designed to detect WebView-induced bugs based on \emph{WebView-induced crashes} and the \emph{lifecycle misalignment criteria} proposed by \wDroid~\cite{hu2018tale}. We identify a crash as WebView-induced if
the crash results from an event executed on an element in the websites loaded by WebViews. 
The lifecycle misalignment criteria were designed to detect UI inconsistencies of a WebView before and after executing certain lifecycle events that make an activity restart (e.g., rotate screen). The assumption is that the displayed web page should be consistent before and after an activity restarts. For instance, the entered information of a form on a web page should not be lost after a device orientation change. Otherwise, the end-user has to re-enter the information in the form.

\subsection{Implementation}
\wTesterws is built on top of Appium~\cite{Appium}, a test automation framework for mobile apps. \wTesterws obtains the UI hierarchy using the UiAutomator2 driver~\cite{uiautomator2} facilitated by Appium. \wTesterws retrieves the covered properties by reading the files that record coverage information dumped by the instrumented AUT (Section~\ref{instru_imple}) through Android's \texttt{adb} debugging tool~\cite{adb}. Following previous work ~\cite{wang2020combodroid,gu2019practical,pan2020reinforcement}, we set a 200 ms delay between events. To switch to other fragment-states (line 14 in Algorithm~\ref{alg:test}), \wTesterws remembers the events that can lead to the switching of fragment-states during testing. One of them (including not-yet-triggered events, events that can open menus/navigation-drawers, etc.) will be picked if \wTesterws decides to switch again. \wTesterws is available online~\cite{wTest}.

\section{Evaluation}\label{evaluation}
In this section, we apply \wTesterws to test real-world Android apps. We evaluate it by studying three research questions:

\begin{itemize}[label=\textbullet, leftmargin=*]
	\item \textbf{RQ1:} Can \wTesterws effectively explore WebView behaviors? Compared with baseline methods, can \wTesterws achieve higher WebView-specific property coverage? 
	 
	
	
	
	\item \textbf{RQ2:} Can \wTesterws effectively detect WebView-induced bugs?
	
	
	\item \textbf{RQ3:} Compared with other coverage criteria, what is the bug-exposing capability of the WebView-specific property coverage criteria?  Is covering more WebView-specific properties helpful to detect more WebView-induced bugs?
	
	
\end{itemize}

\subsection{Evaluation Subjects}
Our evaluation subjects contain \osappsws open-source Android apps and \csappsws closed-source Android apps. They are listed on our website~\cite{wTest}.

\textbf{Collecting open-source apps:}
We collected open-source apps from F-Droid~\cite{FDroid}, a popular open-source app hosting site, and the apps used in \wDroid~\cite{hu2018tale}.
We filtered the apps following the selection criteria below.
An app was excluded if it satisfies one of the following conditions: (1) the app does not use WebViews in its application code; (2) the app does not have any commits within the past three months at the time when we conducted experiments; (3) the app's main functions cannot be reached in a fully automated way (e.g., requires login, remote devices or resources); (4) the app cannot be installed on Android emulators; (5) the app only uses WebViews to display simple pages such as About/License/Advertisement; (6) the app is a toy app (e.g., proof-of-concept apps) or a duplicate of the others.
These criteria enable us to select well-maintained apps with non-trivial use of WebViews.
Following the criteria and excluding the apps that cannot run after instrumentation, \osappsws apps were selected as our experiment subjects. Among them, 32 apps can be found on Google Play. These apps are large-scale ($avg$ over 100 Kloc), well-maintained ($avg$ over 3k revisions),  highly rated ($avg$ 4/5), and diverse (covering 13 categories).

\textbf{Collecting closed-source apps:} We collected closed-source apps using a Google Play crawler Raccoon~\cite{raccoon}. We downloaded the most-popular apps according to the rankings given by AppBrain~\cite{AppBrain1, AppBrain2}. We followed criteria (1), (3), (4), (5) adopted in the previous paragraph when collecting the subjects. We stopped collecting until we successfully instrumented 30 apps. 
These 30 apps cover 15 categories and have over 2.84 billion downloads in total.

\subsection{Experiment Setup}
 \subsubsection{Baselines}
To study the RQs, we selected baseline methods whose tools are publicly available and can run on Android 10. We compared \wTesterws with the following four baselines, including one state-of-the-art WebView test generation technique and three state-of-the-art general-purpose Android test generation techniques.

\begin{itemize}[label=\textbullet, leftmargin=*]
	
	\item \textbf{\wDroid:} \wDroid~\cite{hu2018tale} is a Monkey-based random test generation technique that uses specially designed oracles to detect WebView-induced bugs. The oracles are also adopted by \wTesterws (Section~\ref{oracle}).
	\item \textbf{Q-Testing:} Q-Testing~\cite{pan2020reinforcement} is a reinforcement learning-based general-purpose Android test generation technique. During app exploration, Q-Testing is more likely to pick an event that is expected to obtain higher accumulative rewards (i.e., discover new UI states). Whether two UI states are similar or different is determined by a trained neural network.
	\item \textbf{ComboDroid:} ComboDroid~\cite{wang2020combodroid} is a general-purpose Android test generation technique. Its core idea is to generate a long event sequence (a combo) by combining multiple short event sequences (use cases). ComboDroid tends to combine two use cases where the latter one uses the data written by the former one so as to maximize data-flow diversity. ComboDroid provides a fully-automated variant and a semi-automated variant, and we use the former one. We set the modeling time, a parameter required by ComboDroid, to 30 minutes by following their recommendations.
	
	\item \textbf{Fastbot2:} Fastbot2~\cite{fastbot2} is a general-purpose model-based test generation product from ByteDance~\cite{bytedance}. It builds a probabilistic activity-event transition model and uses reinforcement learning to assist event selection. Fastbot2 uses APE~\cite{gu2019practical} and has been deployed at ByteDance for two years. 
\end{itemize}

\subsubsection{Ablation Study}
To evaluate the necessity of including \wvar s in guiding test generation, we additionally make one more baseline.
\begin{itemize}[label=\textbullet, leftmargin=*]
	
	\item \textbf{\wAPI:} \wAPIws differs from \wTesterws in that it only considers WebView API call sites in the fitness function. This baseline evaluates whether using WebView API call sites alone is adequate to guide WebView testing.
\end{itemize}


\subsubsection{Coverage Calculation}
To measure the effectiveness of exploring WebView behaviors, we
measure the property coverage achieved by \wTesterws and the baselines.
In traditional test generation studies, a coverage percentage (the covered properties over the total number of available properties) is usually used.
However, unlike program statements/methods, the complete set of properties is difficult to obtain in our problem since it requires cross-language data flow analysis that is both sound and complete.
 For fair comparisons, we define the total property set $P_{all}$ for an app as the union of the properties covered by \wTesterws and the baselines when finishing testing.
Then the coverage of a method for an app is defined as:
\begin{equation} \label{cov}
Cov_{method} = \frac{|P_{method}|}{|P_{all}|}
\end{equation}

\subsubsection{Experiment Environment}
We ran experiments on Android emulators 
running Android 10. We choose Android 10 to balance the OS popularity (it is the second most popular Android OS version when we were conducting experiments) and the number of available baselines (e.g., ComboDroid can support up to Android 10). We followed recent works~\cite{hu2018tale,su2017guided, gu2019practical, pan2020reinforcement, sun2021understanding} and allocated one hour to test each app. Since the executions of \wTesterws and all the baselines are subject to some randomness, we repeated the experiments five times to mitigate randomness in the results.
Under these settings, the complete property set of an app was further enlarged to the union of $P_{all}$s over the five rounds. The experiments on closed-source apps were conducted on a machine running CentOS Stream 8, powered by AMD Ryzen Threadripper PRO 3995WX 64-Cores and 512GB memory. The experiments on open-source apps were conducted on a machine running CentOS Stream 8, powered by AMD Ryzen Threadripper 3970X 32-Core Processor and 256GB memory. We ran 16 emulators in parallel on each machine.

\subsection{Results for RQ1 \& RQ2}
In this section, we present the results of WebView-specific property coverage and the detected WebView-induced bugs achieved by each methods. When reporting the coverage results, we will exclude an app for a baseline if it cannot successfully test the app. ComboDroid requires to instrument an app before testing. It fails to instrument 7 open-source apps and 9 closed-source apps in our dataset. Q-Testing cannot successfully test 4 open-source apps and 3 closed-source apps because of multiple engineering issues (e.g., UiAutomator error, app launching failure, etc.). \wDroidws also fails to run on 1 open-source app and 1 closed-source app. We excluded those apps when reporting the coverage results of ComboDroid, Q-Testing, and \wDroid. When reporting the results on the detected bugs, ComboDroid, Q-Testing, and Fastbot2 are not included because they are not equipped with the oracles to detect WebView-induced bugs, and therefore no such bugs can be detected.

\begin{figure}
	\centering
	\includegraphics[width=\linewidth]{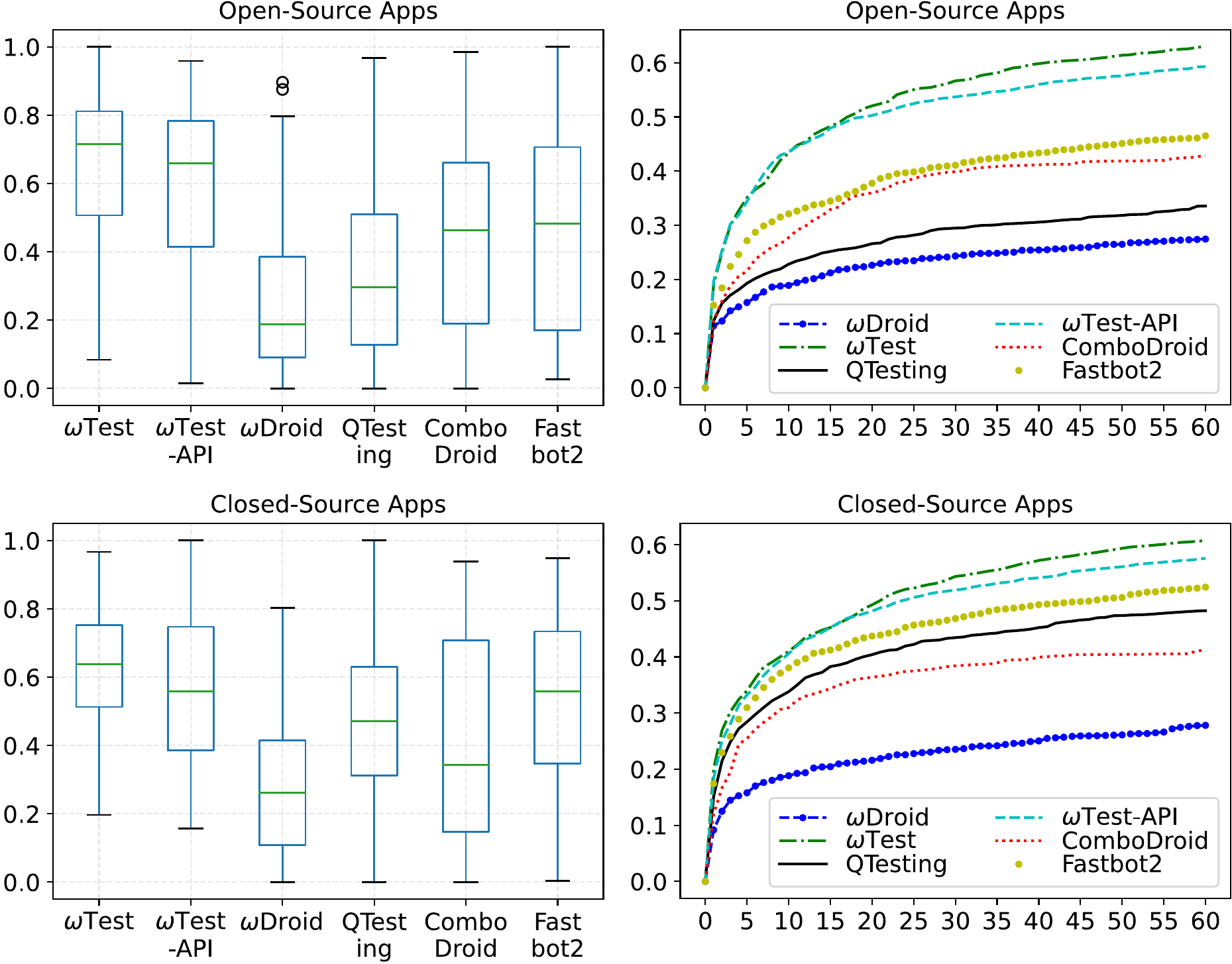}
	\caption{WebView-specific property coverage distributions and its average progressive improvements over 60 mins (The coverage of each app is averaged over 5 experiment rounds)}
	\label{fig:cov}
\end{figure}

\begin{figure}
	\centering
	\includegraphics[width=\linewidth]{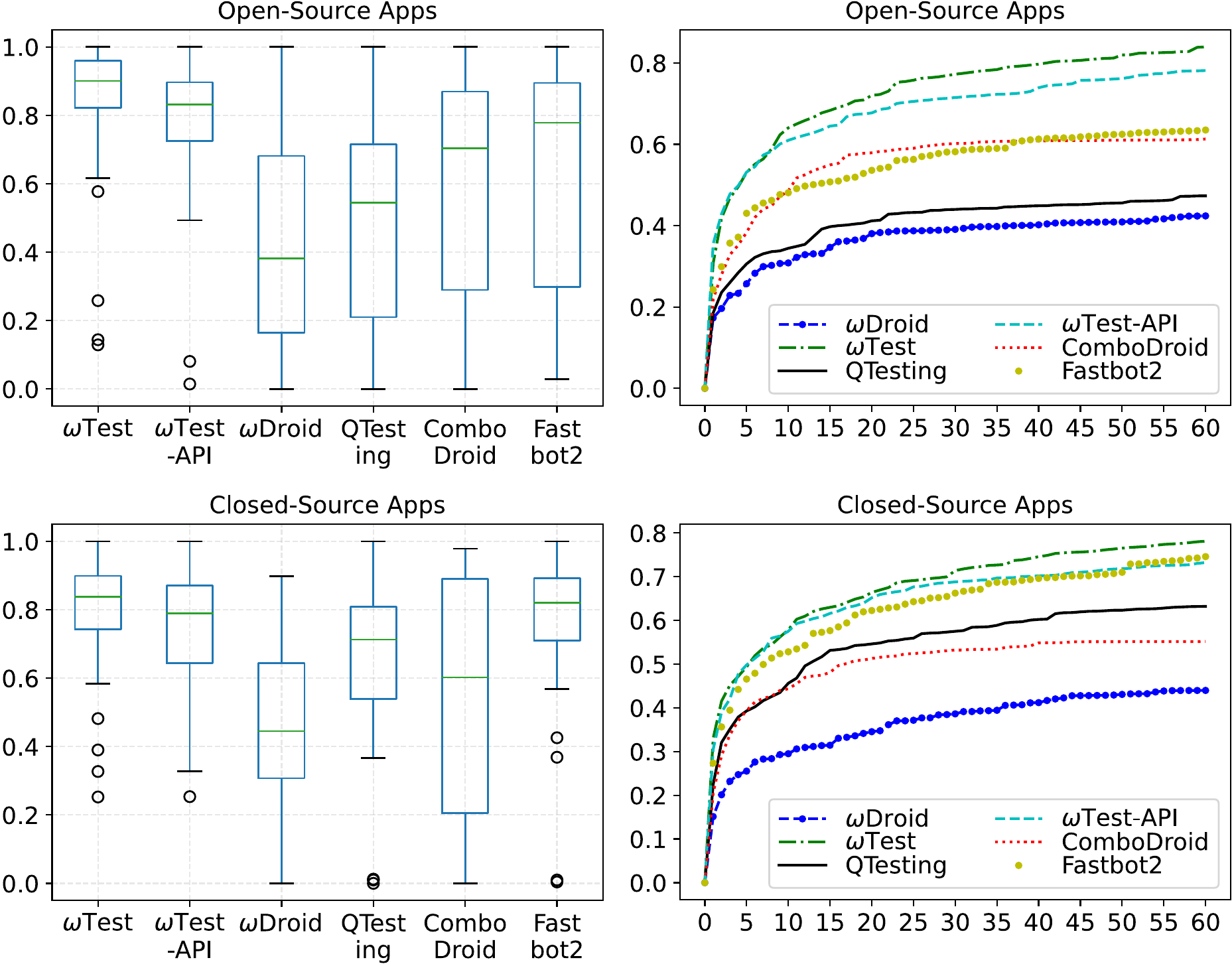}
	\caption{WebView-specific property coverage distributions and its average progressive improvements over 60 mins (The coverage of each app is accumulated from 5 experiment rounds)}
	\label{fig:accucov}
\end{figure}

Figure~\ref{fig:cov} and ~\ref{fig:accucov} illustrate the WebView-specific property coverage distributions and its average progressive improvements of different methods using box plots and line charts. 
In figure~\ref{fig:cov}, each app's coverage achieved by a method is averaged over five rounds. In figure~\ref{fig:accucov}, the properties covered by a method on each app is accumulated from five rounds of experiments.
 The data behind these figures can be found on our website~\cite{wTest}. From the figures, we can see that \wTesterws and \wAPIws can significantly outperform the baseline methods on both open-/closed-source apps. In particular, \wTesterws can achieve 16\%-36\% higher coverage on average on open-source apps and 9\%-33\% higher coverage on average on closed-source apps according to Figure~\ref{fig:cov}. \wTesterws also outperforms \wAPI, which eliminates \wvar s and simply considers WebView API call sites in the fitness function. Compared with \wAPI, \wTesterws can increase the coverage by 4\%(on average)/6\%(on median) and 3\%(on average)/8\%(on median) on open-source apps and closed-source apps, respectively, according to Figure~\ref{fig:cov}.
 Such results indicate that using WebView API call sites alone is inadequate to achieve higher property coverage.

\begin{table}
	\footnotesize
	\begin{center}
		\caption{Number of bugs detected by \wTester, \wAPI, and \wDroidws in 5 rounds of experiments}
		\begin{tabular}{|c|c|c|c|c|c|c|c|c|}
			\ChangeRT{1pt}
			\textbf{App Type} & \textbf{Method} & \textbf{R1} & \textbf{R2} & \textbf{R3} & \textbf{R4} & \textbf{R5} & \textbf{Avg} & \textbf{Total} \\
			\hline	
			\multirow{3}{*}{Open-Source} & \cellcolor{lightgray} \wTester & \cellcolor{lightgray} 16 & \cellcolor{lightgray} 13 & \cellcolor{lightgray} 12 & \cellcolor{lightgray} 12 & \cellcolor{lightgray} 17 & \cellcolor{lightgray} 14 & \cellcolor{lightgray} 24\\
			\cline{2-9}
			& \wAPI & 11 & 11 & 8 & 10 & 12 & 10.4 & 22 \\
            \cline{2-9}
            & \wDroid & 8 & 7 & 7 & 5 & 7 & 6.8 & 11 \\
			\hline
            \ChangeRT{1pt}
            \multirow{3}{*}{Closed-Source} & \cellcolor{lightgray}  \wTester & \cellcolor{lightgray} 8 & \cellcolor{lightgray} 10 & \cellcolor{lightgray} 8 & \cellcolor{lightgray} 6 & \cellcolor{lightgray} 8 & \cellcolor{lightgray} 8 & \cellcolor{lightgray} 12\\
			\cline{2-9}
			& \wAPI & 8 & 5 & 3 & 6 & 5 & 5.4 & 11\\
            \cline{2-9}
            & \wDroid & 3 & 2 & 3 & 2 & 4 & 2.8 & 7\\
			\hline
			\ChangeRT{1pt}
		\end{tabular}
		\label{table:bugs}
	\end{center}
\end{table}

Table~\ref{table:bugs} shows the number of bugs detected by \wTester, \wAPI, and \wDroidws over the five rounds of experiments (ComboDroid, Q-Testing, and Fastbot2 do not have the oracles to detected WebView-induced bugs). In total, these 3 methods detected 27 bugs in 19 open-source apps and 13 bugs in 8 closed-source apps in these 5 runs. On average, \wTesterws detected more number of bugs than \wAPIws and \wDroid.
\wTesterws failed to find three bugs that were detected by \wDroidws in open-source apps because specific event types are not supported by \wTesterws currently. \wTesterws failed to find one bug that was detected by \wDroidws and \wAPIws in a closed-source app because we found the fitness value drops quickly on the activity that contains the buggy WebView in that app. Therefore \wTesterws decides to spend fewer events on that activity, leaving the bug undetected. 
We reported all the bugs detected in open-source apps to their corresponding GitHub issue trackers and provide the issue links on our website~\cite{wTest}. To comply with each app's contributing guide and license, we thoroughly tested each app and submitted a bug report in a proper format if the bug  can be consistently reproduced. 
If a GitHub repository is archived or the detected bug has already been fixed in newer versions of the app, we provide the reproducing steps on our website~\cite{wTest}. We also provide the reproducing steps for bugs detected in closed-source apps on our website~\cite{wTest}. Among the 22 submitted bug reports whose bugs were detected by \wTester, \confirmedBugsws  bugs have been confirmed and \fixedBugsws of them have been fixed.

During bug reproduction, we observed that covering \wvar s is helpful in driving \wTesterws to explore diverse WebView behaviors, thus discovering more hidden bugs in different usage scenarios of WebViews. For example, \wTesterws consistently detected bugs in Notepad~\cite{NotepadFree}, which is a notes edit app that has over 10 million downloads. The app uses a WebView to display Frequently Ask Questions (FAQs). \wTesterws discovered that a user's reading progress would get lost after device rotation because the WebView refreshes the page after rotation. In another scenario, Notepad uses a WebView to display Privacy Policies (PPs). Exploring FAQs and PPs will cover the same set of WebView API call sites but different sets of \wvar s. \wAPIws, which is solely driven by WebView API call sites, may consider these scenarios being adequately explored after a few events, thus missing the bug hidden in these scenarios.

In summary, \wTesterws can effectively explore WebView behaviors of Android apps. It can achieve higher WebView-specific property coverage and detect more number of WebView-induced bugs.

\begin{figure}
	\centering
	\includegraphics[width=\linewidth]{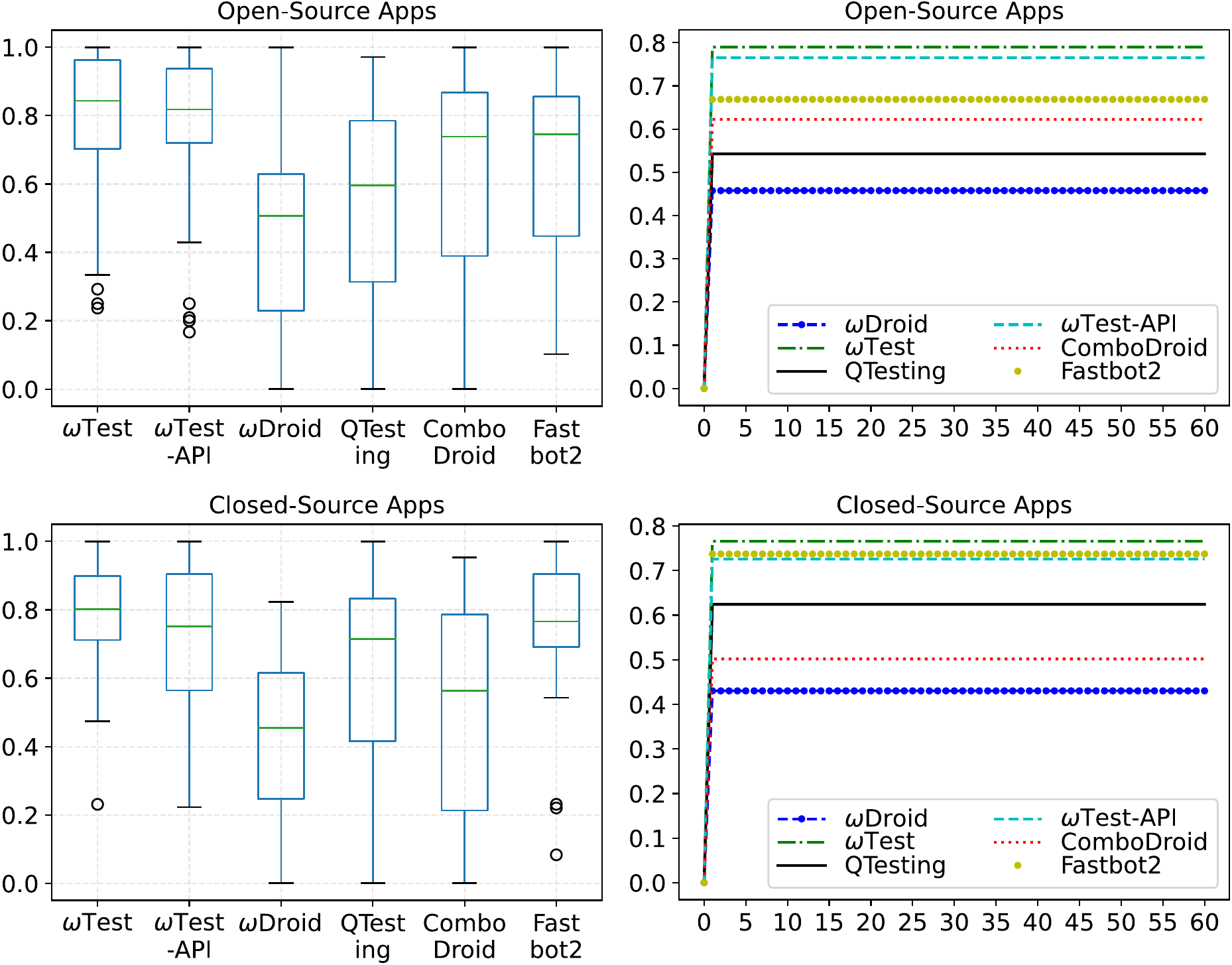}
	\caption{WebView API call site coverage distributions and its average progressive improvements over 60 mins (The coverage of each app is averaged over 5 rounds of experiments)}
	\label{fig:wapiCov}
\end{figure}

\begin{figure}
	\centering
	\includegraphics[width=\linewidth]{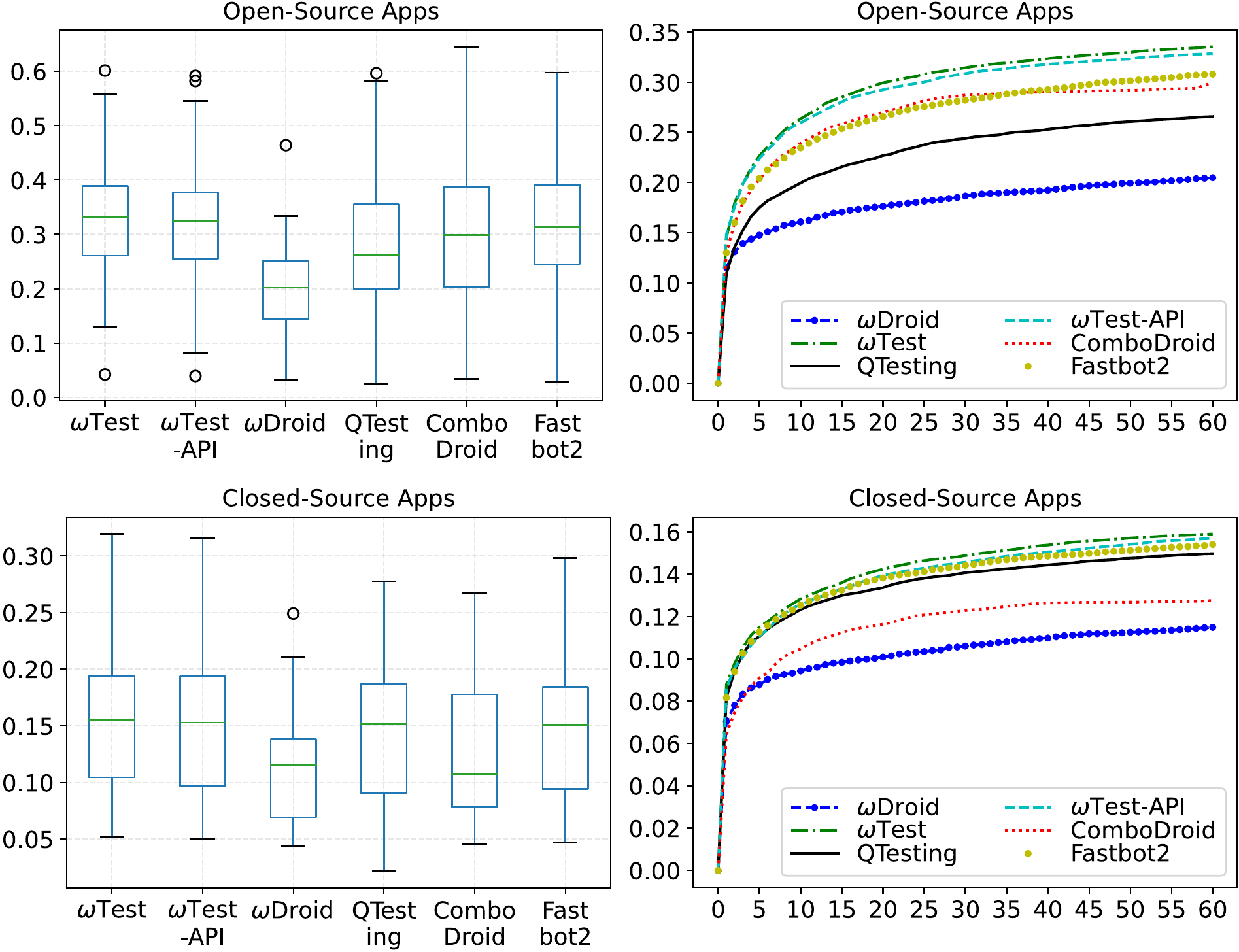}
	\caption{Code coverage distributions and its average progressive improvements over 60 mins (The coverage of each app is averaged over 5 rounds of experiments)}
	\label{fig:codeCov}
\end{figure}

\subsection{Results for RQ3}

\begin{figure}
	\centering
	\includegraphics[width=\linewidth]{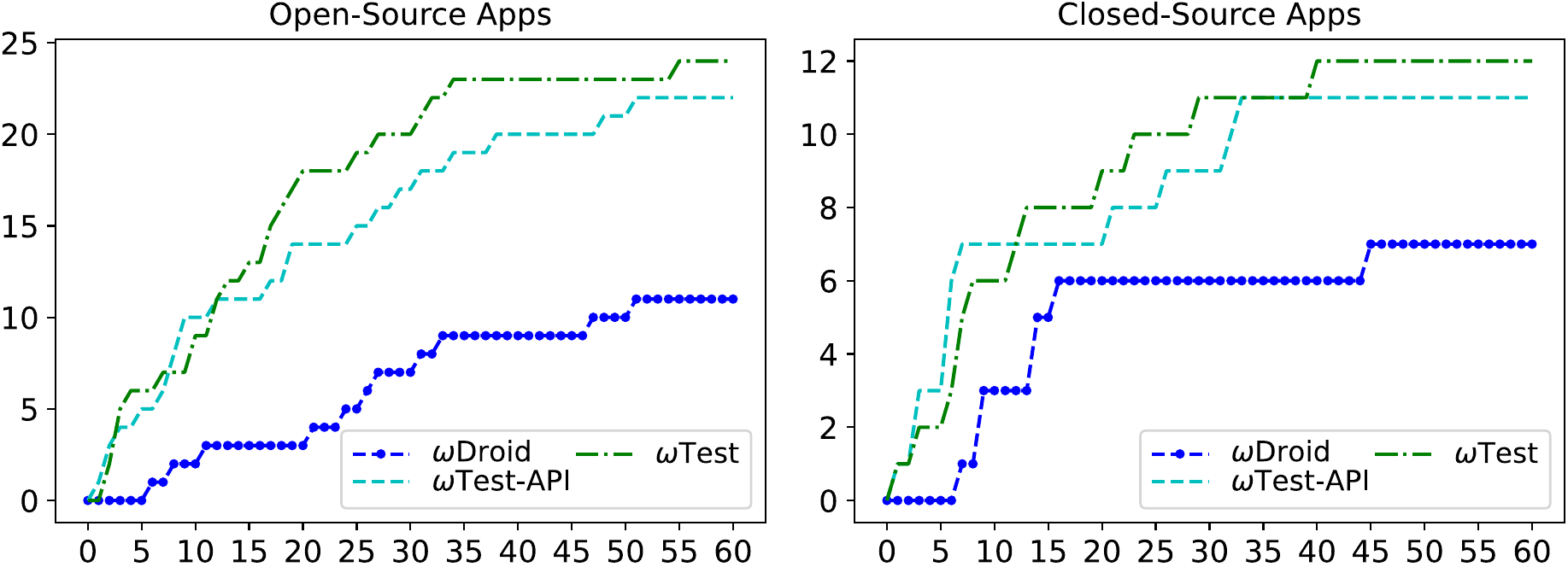}
	\caption{Progressive improvements of the detected bugs over 60 mins}
	\label{fig:bugCounts}
\end{figure}

To demonstrate the bug-exposing capability of the WebView-specific property coverage criteria, we compare it with WebView API call site coverage criteria and code coverage criteria. Figure~\ref{fig:wapiCov}-\ref{fig:codeCov} show the coverage distributions and the progressive improvements of WebView API call cite coverage and code coverage, respectively. In addition to the number of bugs detected at the end of testing shown in Table~\ref{table:bugs}, we plot the progressive improvements of the number of bugs detected by each method in Figure~\ref{fig:bugCounts}. The time that a bug is found by a method is the first time when it is detected. The time is averaged by a number between 1 to 5, depending on how many times that the bug is detected in the five rounds of experiments. 

Figures~\ref{fig:wapiCov} and \ref{fig:bugCounts} suggest that the coverage criterion based on WebView API call sites has a weak correlation with the number of detected bugs. The coverage quickly converges at the initial stage of the testing (before 5 minutes). However, many bugs are detected after 5 minutes. 
In addition, Figure~\ref{fig:codeCov} shows the code coverage achieved by \wTesterws and \wAPIws are similar on both open-source apps (avg 33.5\% vs 32.9\%) and closed-source apps (avg 15.9\% vs 15.7\%). However, Table~\ref{table:bugs} shows that \wTesterws detects far more bugs than \wAPI, which sugguests that code coverage criterion also has a weak correlation with the number of detected bugs. 

To quantitatively measure the correlations, we follow existing work~\cite{wang2019map, inozemtseva2014coverage, kochhar2017code} to compute Kendall correlations~\cite{kendall1938new} between the coverage computed base on a criterion and the number of detected bugs. Kendall correlation measures the correlations between two sets of data. A value between 0 to 1 indicates they are positively correlated (0-0.4 means a low correlation, 0.4-0.7 means a moderate correlation, 0.7-1 means a strong correlation). We compute the correlations between the average coverage on the buggy apps and the number of dectected bugs achieved by each method in each round at 5,10,15,...,60 minutes. The results for WebView-specific property coverage criteria, WebView API call site coverage criteria, and code coverage criteria on open-source apps are 0.7, 0.49, and 0.67, respectively. The results for WebView-specific property coverage criteria and WebView API call site coverage criteria on closed-source apps are 0.59 and 0.53, respectively. The result on code coverage is not reliable becasue closed-source apps are highly obfuscated. The code coverage are severely affected by third-party/system libraries, which cannot be effectively distinguished from application code. The results suggest that WebView-specific property coverage criteria has a moderate to strong correlations with the number of detected bugs. 

In summary, the WebView-specific property coverage criteria have a stronger bug-exposing capability than the WebView API call site coverage criterion and the code coverage criterion. Covering more WebView-specific properties is helpful in detecting more WebView-induced bugs.

\subsection{Limitations}
We observed a limitation of \wTesterws when analyzing the experiment results. We found \wTesterws is not effective in reaching difficult-to-reach WebViews in an Android app. Some WebViews are deeply hidden in an app (e.g., a long activity stack is observed when the WebView is reached) so that \wTesterws may decide to leave a component too early, but actually more events deserve to be tried. Furthermore, we also observed that some WebViews cannot be reached until specific conditions are satisfied. For example, a weather app called RadarWeather~\cite{RadarWeather} uses a WebView to display a weather map in an activity. The map is only available until users enters a city name in another activity. The fitness function adopted by \wTesterws is not able to model such information. In future work, we plan to extend \wTesterws with static analysis, which aims to identify the activities/fragments that are helpful to reach WebViews.

\subsection{Threats to Validity}
Our property coverage (Formula~\ref{cov}) may not reflect the ``real'' coverage since $P_{all}$ can be different from the complete set of WebView-specific properties (there can be properties that were not discovered in our experiments). Obtaining the complete set in our problem is difficult because it requires cross-language data flow analysis that is both sound and complete. It also requires a sound and complete string analysis to predict the possible JavaScript code that is dynamically constructed at runtime.
We mitigated this problem by approximating the complete set with the union of the properties identified by the six methods over five runs.
Although our coverage results can be different from that computed using the complete property set, it provides a reliable evaluation of the relative performance achieved by different methods. This meets our evaluation goals and it is adopted by existing work~\cite{wang2019map,terragni-icse-2016} in which the complete set is hard to obtain.

The conclusions drawn from our evaluation are affected by the representativeness of the selected subjects. To mitigate the threat, we selected \osappsws real-world open-source Android apps that are large-scale, well-maintained, and diverse in categories. We also included \csappsws closed-source apps that are selected from the most popular Android apps on the Google Play store. They have more than 2.8 billion downloads in total.

Randomness can affect our evaluation results as the algorithms of \wTesterws and all the baselines involve randomness. To mitigate the issue, we follow existing works~\cite{su2017guided, gu2019practical, dong2020time} to repeat our experiments five times.

\subsection{Discussion}
The WebView-specific properties are computed via a set of propagation rules based on explicit data flows. Like existing work~\cite{bell2014phosphor,bell2015dynamic}, we choose not to propagate \wvar s through implicit operations (e.g., control flows) to reduce noise. Except \wvar s, coverage could be measured based on other types of properties such as call chains that involve WebView API calls, which seems sufficient for revealing the bugs. However, we choose not to use such call chains because (1) determining calls relevant to WebViews is difficult. Appending all calls around WebView API calls can include irrelevant methods. In comparison, the analysis defined in Section~\ref{property} can effectively determine the part of an app that is relevant to WebViews. (2) Determining the length of call chains is hard. Shorter chains may have a weak bug detection capability. For example, the covered calls for cs-bug1 and cs-bug2~\cite{wTest} in Notepad~\cite{NotepadFree} are the same if the length of the chain is smaller than 100 (50 calls before and after WebView API calls). Longer chains may increase the bug-detection capability, but can include many methods irrelevant to WebViews, complicating the analysis and increasing test cost.
\section{Related Work}\label{related}
\subsection{WebView Study}
WebView has attracted immense interest from research communities over the past ten years mainly because of the new security threats it brings to Android apps. 
Many attack models and mitigation solutions were proposed. For example, 
Bai et al. proposed BridgeTaint~\cite{bai2018bridgetaint}, a dynamic taint tracking technique targeting the WebView's bridge communication~\cite{webApps} to detect privacy leaks and cross-language code injection attacks. More recently, Yang et al. proposed EOEDroid~\cite{yang2018automated}, OSV-Hunter~\cite{yang2018study}, and DCV-Hunter~\cite{yang2019iframes} that detect three kinds of new vulnerabilities resulting from WebView's event handlers, postMessage mechanism, and iframes/popups, respectively. Hu et al. proposed \wDroid~\cite{hu2018tale} to detect WebView-induced lifecycle misalignment bugs. Our work complements existing work because we focus on effective test generation to examine WebView behaviors in an Android app. 


\subsection{Android Testing}
A large number of test generation techniques have been proposed to test Android apps~\cite{choudhary2015automated,wang2018empirical,su2021benchmarking}. 
They can be classified into two major categories according to their test objectives. One is model-based test generation~\cite{su2017guided, baek2016automated, gu2019practical, pan2020reinforcement, dong2020time, degott2019learning, fastbot2} whose objective is to discover diverse GUI states of an Android app. 
Intuitively, more discovered GUI states that ``look different'' means more app behaviors are explored. Another major category looks for program properties (e.g., program statements and branches) in an app's program structure and takes them as the test objectives~\cite{su2017guided,mao2016sapienz, mahmood2014evodroid,wang2020combodroid}. 
Although these existing program properties may be suitable for general-purpose testing, none of them are designed for testing WebViews in Android apps. In our paper, we design a novel coverage criterion based on WebView-specific properties to guide the test generation to effectively examine WebView behaviors in Android apps.
\section{Conclusion}\label{conclusion}
In this paper, we proposed a novel design of WebView-specific properties that can abstract WebView behaviors in Android apps.
The property can be utilized to guide test generation to explore diverse WebView behaviors.
Based on this idea, we devised \wTester, a test generation technique that maximizes the number of covered WebView-specific properties.
Our evaluation results show that \wTesterws can effectively generate tests exercising diverse WebView behaviors and detect WebView-induced bugs.
\wTesterws now only adopts crashing and lifecycle misalignment oracles. In the future, we plan to extend the oracles and leverage \wTesterws to detect more types of  WebView-induced bugs. We will also study how to effectively reach difficult-to-reach WebViews in our future work.
\section{Data Availability}\label{data}
We make all our data publicly available at \url{https://richardhooooo.github.io/wTest/}. The website includes (1) open-source and closed-source apps used in experiments, (2) coverage results achieved by \wTesterws and the baselines, (2) the links to the submitted bug reports and reproduction steps if the report is not able to be submitted, (3) the customized Android OS, (4) the tool \wTesterws and its guidance.

\begin{acks}
The authors thank ISSTA 2023 reviewers. This work is supported by National Natural Science Foundation of China (Grant No. 61932021), Hong Kong Research Grant Council/General Research Fund (Grant No. 16211919), Hong Kong Research Grant Council/Research Impact Fund (Grant No. R503418), NSERC Discovery Grant RGPIN-2022-03744 DGECR-2022-00378, and Guangdong Basic and Applied Basic Research Fund (Grant No. 2021A1515011562).
\end{acks}

\balance
\bibliographystyle{ACM-Reference-Format}
\bibliography{reference}


\end{document}